\documentclass[aps,prb,amsmath,amssymb,superscriptaddress,floatfix,showpacs,preprint]{revtex4}

\usepackage{graphicx}
\usepackage{bm}
\input epsf.sty

\begin{document}


\title{Aging in coarsening diluted ferromagnets}

\author{Hyunhang Park and Michel Pleimling}
\affiliation{Department of Physics, Virginia Polytechnic Institute and State University, Blacksburg, Virginia 24061-0435, USA}

\date{\today}

\begin{abstract}
We comprehensively study non-equilibrium relaxation and aging processes in the two-dimensional random-site
Ising model through numerical simulations. We discuss the dynamical correlation length as well as
scaling functions of various two-times quantities as a function of temperature and of the 
degree of dilution. For already modest values of the dynamical correlation length $L$ deviations from a
simple algebraic growth, $L(t) \sim t^{1/z}$, are observed. 
When taking this non-algebraic growth properly into account, a simple aging behavior of the autocorrelation
function is found. This is in stark contrast to earlier studies where, based on the assumption of algebraic 
growth, a superaging scenario was postulated for the autocorrelation function in disordered ferromagnets.
We also study the
scaling behavior of the space-time correlation as well as of the time integrated linear response and 
find again agreement with simple aging. Finally, we briefly discuss
the possibility of superuniversality in the scaling properties of space- and time-dependent quantities.
\end{abstract}


\maketitle
\pagestyle{plain}
\section{Introduction}
Understanding the effect of randomness and disorder remains one of the main
challenges in condensed matter physics and materials science. One issue that has attracted much
attention recently concerns ordering processes, relaxation phenomena, and the formation of domains in disordered systems. 
Examples include vortex lines in disordered type-II superconductors,\cite{Nic02,Ols03,Bus06,Bus07,Du07}  polymers in random 
media,\cite{Kol05,Noh09,Igu09,Mon09} coarsening of
disordered magnets,\cite{Pau04,Pau05,Pau07,Hen08,Aro08,Lou10} and slow dynamics in spin glasses.\cite{Aro08,Sib10,Fer10}

Our understanding of relaxation phenomena and aging processes in non-frustrated systems with slow dynamics
has greatly progressed during the last decade, mainly through the systematic study of critical
and coarsening ferromagnets (see [\onlinecite{HenPle}] for a review of the field). Whereas aging in perfect, i.e.
non-disordered, systems is now very well understood, this is different for disordered systems where almost 
every aspect of non-equilibrium relaxation remains under debate. 

Already the growth law governing the coarsening process of disordered ferromagnets below their critical temperatures
is at the center of some controversy. Monte Carlo simulations
of various disordered ferromagnetic Ising models \cite{Pau04,Pau05,Pau07,Hen08,Aro08,Lou10} found for
the dynamical correlation length $L(t)$ a power-law increase,
$L(t) \sim t^{1/z}$, with a non-universal dynamical exponent $z$ that depends on temperature and on the nature
of the disorder. This behavior can be explained by assuming that the energy barriers grow
logarithmically with $L$.\cite{Pau04,Pau05} As a consequence,
it follows that the dynamical exponent $z$ should be inversely proportional to the temperature: $z \sim 1/T$.
Whereas this temperature dependence has been observed in the Cardy-Ostlund model,\cite{Sch05} a careful study \cite{Hen08}
has revealed that in the random-bond Ising model the temperature dependence of the dynamical exponent
is more complicated. This fact points to the existence of conceptual problems in the approach of Paul {\it et al}.\cite{Pau04,Pau05}
In addition, an algebraic growth-law is in strong contrast to the classical theory of activated dynamics that,
under the assumption of energy barriers growing as a power of $L$, predicts a slow
logarithmic increase \cite{Hus85} of this length: $L \sim  (\ln t )^{1/\psi}$, with the barrier exponent $\psi > 0$.
New insights into this matter have come very recently through a series of papers on the dynamics of elastic lines
in a random potential.\cite{Noh09,Igu09,Mon09} These papers provide convincing evidence for a dynamic crossover
between a transient regime, characterized by a power-law growth with an effective dynamical exponent that depends on the
disorder, and the asymptotic regime where the growth is logarithmic in time.
This result strongly suggests the possibility that a similar crossover could take place in disordered
ferromagnets and that the observed power-law regime with non-universal dynamical exponents is 
not the asymptotic regime.
However, as the transient regime is already extremely long-lived for the elastic line,\cite{Noh09,Igu09}
it is expected that the asymptotic regime is not easily accessible in simulations of coarsening ferromagnets.
Nevertheless, it can well be that on the time scales accessible in relaxation
measurements a crossover takes place.

Another question concerns the scaling behavior of two-times quantities encountered in disordered ferromagnets 
relaxing towards equilibrium.
In many systems it is empirically observed that in the aging regime the autocorrelation function $C(t,s)$,
where $s$ is the waiting time and $t > s$ is the observation time, behaves like \cite{HenPle}
\begin{equation}
C(t,s) = {\cal C} \left(\frac{h(t)}{h(s)}\right) \;\; , \;\;
h(t) = h_0 \exp\left[ \frac{1}{c_0}\frac{t^{1-\mu}-1}{1-\mu}\right]
\label{eq:gen_scal}
\end{equation}
where ${\cal C}$ is a scaling function, $\mu$ is a free parameter and
$h_0$ and $c_0$ are constants.\cite{And06}  Depending on the value of $\mu$, 
different situations are summarized by Eq. (\ref{eq:gen_scal}).
If $0 < \mu < 1$, a subaging behavior prevails, as is for example observed in soft matter \cite{Ram01,Wan06}
or in living biological matter,\cite{Tre07} 
with $\mu \approx 0.3 - 0.8$. For $\mu =1$, one recovers a standard simple aging
behavior where the autocorrelation is only a function of the ratio $t/s$. This is the behavior encountered in many systems,
and in particular in the perfect
ferromagnets undergoing phase ordering. Finally, for $\mu >1$ one has a superaging behavior. This is a rather hypothetical
scenario as Kurchan has proven an exact lemma \cite{Kur02} that states that under very general conditions a superaging behavior
of the autocorrelation can not exist (this prove can also be extended to response functions \cite{HenPle}). 
It is therefore intriguing that a study of the autocorrelation in 
the two-dimensional random-site Ising model \cite{Pau07} yielded data 
to which one can fit the superaging scaling form 
\begin{equation} \label{C_super}
C(t,s) = {\cal C}\bigl( \exp\bigl[ \frac{(t-s)^{1-\mu}-s^{1-\mu}}{1-\mu}\bigr]\bigr)
\end{equation}
with $\mu$ slightly larger than 1. However, in the random-bond Ising model 
the autocorrelation displays a simple aging behavior \cite{Hen08}
in full analogy to what is observed in the pure model.\cite{foot0}
These differences in the observed scaling behavior of different disordered ferromagnets remains to be explained.

Finally, some coarsening ferromagnets with weak randomness \cite{Pur04} 
(as for example the random-bond or the 
random-field Ising models) have been shown to display a {\it superuniversal} behavior: scaling functions of space- and time-dependent
quantities are independent of disorder and temperature provided that distances are measured in units of the dynamical
correlation length $L(t)$.\cite{Fis88} This has been verified in various coarsening systems
both for one-time \cite{Bra91,Pur91,Hay91,Bis96,Aro08,Lou10} and two-times quantities,\cite{Hen08} 
but it is an open question how general
this result really is.

In this paper we are revisiting the random-site Ising model in two dimensions. Our simulations reveal that for this system
the apparent power-law growth of the dynamical correlation length $L(t)$ only holds in the early stages of the coarsening process
and that notable deviations are already encountered for moderate values of the correlation length. As a consequence one can not
naively assume that $L(t) \sim t^{1/z}$, as done in previous studies,\cite{Pau05,Pau07}  but instead the correct time dependence
$L(t)$ has to be used in the investigation of the scaling behavior of two-times quantities. 
When doing this, a simple aging behavior emerges for the autocorrelation function
in the scaling limit, thereby showing that the previously proposed superaging scenario is not appropriate for
the description of aging in diluted ferromagnets relaxing towards equilibrium.

We extend our study to the time integrated linear response and to the 
space-time correlation function and find that both quantities also show a simple aging scaling.
Finally, we also investigate a possible superuniversal
behavior in diluted ferromagnets 
and show that the space-time correlation functions
for a fixed degree of dilution but different temperatures fall approximately on a common master curve.
This is in contrast to data obtained for different dilutions for which a data collapse is not observed.

Our paper is organized in the following way. In the next Section we introduce the model and the quantities (autocorrelation,
space-time correlation, autoresponse) that we study in order to elucidate the aging and dynamical scaling properties of
diluted ferromagnets. In Section III we present the results from our extensive numerical simulations. As a function of dilution
and temperature we determine the dynamical growth law as well as non-equilibrium exponents and establish the scaling behavior of the
various quantities. Finally, Section IV gives our conclusions.

\section{Model and measured quantities}
We consider in the following the two-dimensional random-site Ising model \cite{Sta79} on a square lattice (we set the 
lattice constant to one) with the
Hamiltonian
\begin{equation}
{\mathcal H} = - \sum\limits_{\langle \bf{x},\bf{y} \rangle} \rho_{\bf{x}} \rho_{\bf{y}} 
S_{\bf{x}} S_{\bf{y}}
\end{equation}
where the sum is over nearest neighbor pairs. $S_{\bf{x}} = \pm 1$ are the usual Ising spins
and $\rho_{\bf{x}}$ are quenched random variables taken from the distribution $P(\rho) = p \delta_{\rho,1} + (1-p) 
\delta_{\rho,0}$. For $p=1$ we of course recover the standard Ising model on a square lattice. If the system 
is diluted, i.e. if $p < 1$, one still observes a phase transition between a disordered high temperature phase and an
ordered low temperature phase for all $p \geq p_c \approx 0.593$. 

In our simulations we prepare the system in a fully disordered state, corresponding to infinite temperature,
and then bring it in contact with a heat bath at a temperature $T < T_c(p)$ where $T_c(p)$ is the critical temperature
for the dilution $p$. After the quench, the system is evolved with the standard single spin flip Metropolis
algorithm (in order to check that our conclusions are independent of the update scheme we also did a large number
of runs with the heat-bath algorithm),
and the relaxation of the system towards equilibrium is monitored through the study 
of the autocorrelation, the space-time correlation and the thermoremanent
susceptibility. The two-time autocorrelation function is given by the expression
\begin{equation}
C(t,s) = \frac{1}{N} \sum\limits_{\bf{x}} \overline{\langle S_{\bf{x}}(t) S_{\bf{x}}(s) \rangle}
\end{equation}
where the sum is over the $N = L^2$ lattice sites and $S_{\bf{x}}(t)$ is the value of the spin at site
$\bf{x}$ at time $t$. For all studied quantities we have to average over both the thermal noise
(indicated by $\langle \cdots  \rangle$) and the site disorder (indicated by $\overline{\cdots}$). In order to get
a more complete picture of the relaxation processes, we extend our 
study to the two-time space-time correlation function
\begin{equation}
C(t,s;{\bf{r}}) = \frac{1}{N} \sum\limits_{\bf{x}} 
\overline{\langle S_{\bf{x}+\bf{r}}(t) S_{\bf{x}}(s) \rangle}~.
\end{equation}
Finally, we also investigate how the system reacts to a perturbation by adding a small spatially random magnetic 
field $h_{\bf{x}} = H_0 \varepsilon_{\bf{x}}$, $\varepsilon_{\bf{x}} = \pm 1$, at the moment of the quench.\cite{Bar98}
This field is turned off after the waiting time $s$ after which we measure the decay of the time-dependent thermoremanent 
susceptibility 
\begin{equation} \label{eq:trm}
\chi(t,s) = \frac{1}{N} \sum\limits_{\bf{x}} \left\{ \overline{ \langle \varepsilon_{\bf{x}} S_{\bf{x}} \rangle } \right\}~~, ~ t> s~,
\end{equation}
where an additional average over the random field, indicated by $\left\{ \cdots \right\}$, has to be performed.

The results reported in the next Section have been obtained for systems composed of $N = 300 \times 300$ spins.
We carefully checked that no finite-size effects show up in our simulations for this system size. We considered three different degrees
of dilution: $p = 0.9$, 0.8, 0.75, and three different temperatures for every $p$ value: $T= 0.7 \, T_c(p)$, 
$0.5 \, T_c(p)$, $0.4 \, T_c(p)$, with $T_c(0.9) \approx 1.914$, $T_c(0.8) \approx 1.50$, $T_c(0.75) \approx 1.30$.
For the autocorrelation and the space-time correlation we averaged typically over at least 1000 independent runs, thereby
considering waiting times $s$ up to 20000 MCS, with a total running time of $t = 50 s$ MCS
(as usual, time is measured in Monte Carlo Steps, one MCS corresponding to $N = L^2$ proposed updates of the system). 
We also studied the corresponding quantities with $s=0$, i.e. $C(t,s=0)$ and
$C(t,s=0,{\bf{r}})$, as well as the one-time quantity $C(t, {\bf{r}}) \equiv C(t,t,{\bf{r}})$. For the autoresponse
we used $H_0=0.05$ (we checked that we are in the linear response regime for that strength of the random field) and
averaged at least over 10000 independent runs. We thereby considered waiting times up to $s= 800$ and observation times
again up to $t = 50 \, s$.

Before closing this Section, we briefly summarize for later the scaling forms expected for our quantities in case of a simple
aging scenario. In the aging regime we have that $t$, $s$, and the difference $t-s$ are large compared to any
microscopic time scale. If simple aging prevails, one expects that the autocorrelation scales as
\begin{equation} \label{eq:Cscal}
C(t,s) = s^{-b} f_C(t/s)
\end{equation}
where $b$ is a non-equilibrium exponent and $f_C(y)$ is a scaling function which displays a power-law in the 
long-time limit: $f_C(y) \sim y^{-\lambda_C/z}$ for $y \gg 1$. Here $\lambda_C$ is the autocorrelation exponent \cite{Fis88}
and $z$ is the dynamical exponent. For systems undergoing phase-ordering (as for example the perfect
kinetic Ising model) it is usually found that $b=0$, see [\onlinecite{HenPle}], but some other classes of systems, for example non-equilibrium growth
systems,\cite{Rot06,Cho10}  are known to have $b \neq 0$. 

It is important to note that the scaling form (\ref{eq:Cscal}) {\it assumes} a power-law growth law of the dynamical correlation
length: $L(t) \sim t^{1/z}$. However, as we shall see, deviations from a simple power-law growth of the dynamical correlation length
already show up for moderate values of $L$. A more general scaling form in terms of the correlation length,
valid in the scaling regime $1 \ll L(s) \ll L(t)$, is given by
\begin{equation} \label{eq:Cscal2}
C(t,s) = \left(L(s)\right)^{-B} \widetilde{f}_C\left(\frac{L(t)}{L(s)}\right)
\end{equation}
with $\widetilde{f}_C(y) \sim y^{-\lambda_C}$ for $y \gg 1$.  

For the space-time correlation function one then expects the following scaling behavior:
\begin{equation} \label{eq:Cscal3}
C(t,s,{\bf{r}}) = \left(L(s)\right)^{-B} \widetilde{F}_C\left( \frac{L(t)}{L(s)}, \frac{\left| \bf{r} \right|}{L(t)} \right) 
\end{equation}
where $\widetilde{F}_C\left( y,z \right)$ is a space- and time-dependent scaling function. In case of a power-law increase of $L$,
the length $L(t)$ resp. $L(s)$ can be replaced by $t^{1/z}$ resp. $s^{1/z}$, yielding the 
relation $b = B/z$. Finally, for the thermoremanent susceptibility,
which as a time integrated response is related to the autoresponse $R(t,s)$ by 
$\chi(t,s) = \int\limits_{0}^s \, R(t,u) \, du$, one should have
that
\begin{equation} \label{eq:trm2}
\chi(t,s) = \left(L(s)\right)^{-A} \widetilde{f}_M\left(L(t)/L(s)\right)
\end{equation}
in the dynamical scaling regime. The scaling function $\widetilde{f}_{\chi}(y)$ is asymptotically given by a power-law decay,
$\widetilde{f}_{\chi}(y) \sim y^{-\lambda_R}$ for $y \gg 1$, with the autoresponse exponent $\lambda_R$.\cite{Pic02} Again,
in case $L(t) \sim t^{1/z}$, one can write this in the widely used form
\begin{equation} \label{eq:trm3}
\chi(t,s) = s^{-a} f_{\chi}(t/s)
\end{equation}
where $a = A/z$ and $f_{\chi}(y)  \sim y^{-\lambda_R/z}$ for $y \gg 1$.
In the perfect two-dimensional Ising model, one has that $a = 1/z$.\cite{HenPle} 
This is different for the random-bond Ising model, as here one finds that $a \neq 1/z$.\cite{Hen08,Lip10}

\section{Numerical results}
In the following we first discuss the time dependence of the correlation length before studying in a systematic way
the scaling behavior of the various two-times quantities introduced in the previous Section.

\subsection{Dynamical correlation length}
\begin{figure}[h]
\centerline{\epsfxsize=4.25in\ \epsfbox{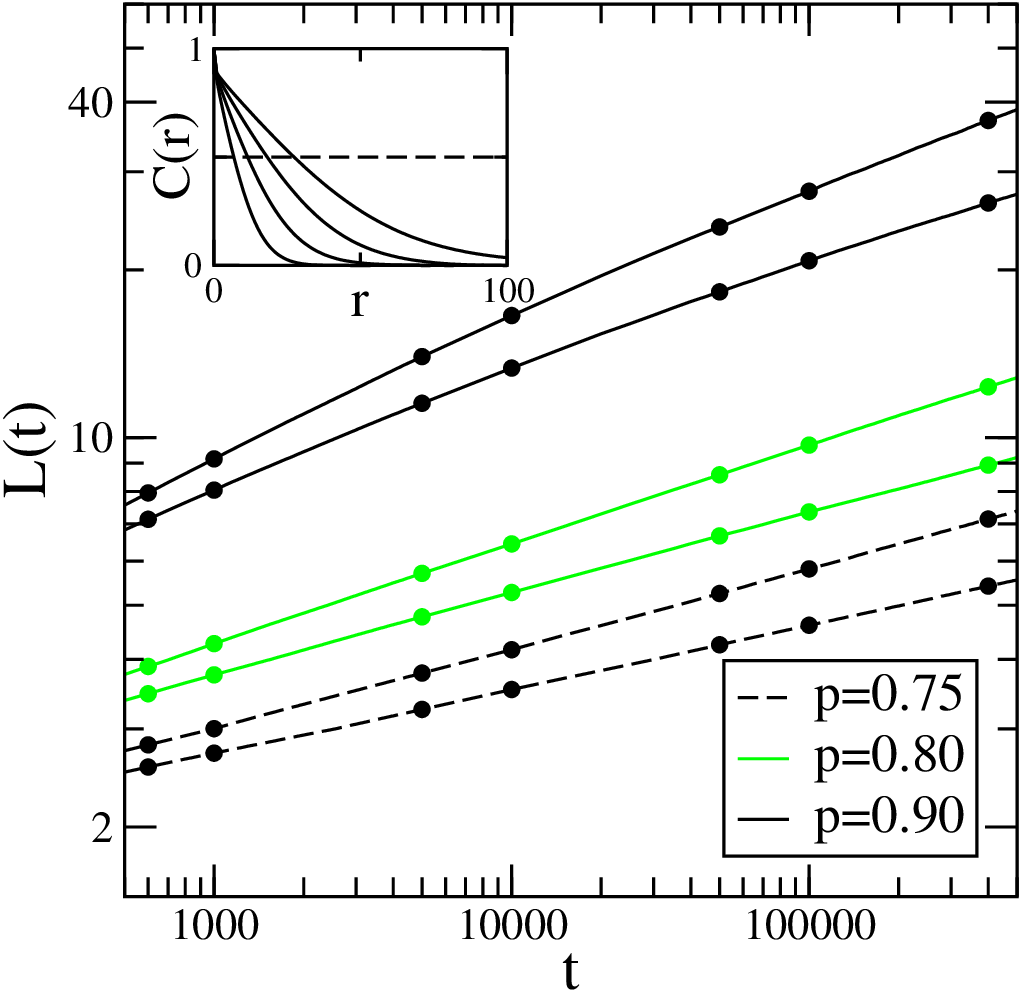}}
\caption{[Color online] Dynamical correlation length $L(t)$ {\it vs} time. For every degree of dilution, data for two temperatures
are shown: $T= 0.4 \, T_C(p)$ and $0.5 \, T_C(p)$ (from bottom to top). After an early time regime where $L(t)$ is
effectively described by a power-law in time, deviations from a simple algebraic growth start to be manifest already for moderate 
values of $L(t)$. For some selected times we represent the data by dots. The inset shows for $p = 0.90$
and $T = 0.4 \, T_C(0.90)$ the decay of the spatial correlation function for $t = 500$, 5000, 50000, and 500000 (from left
to right). $L(t)$ is obtained from these curves by their intersections with the dashed line $C(r) = 0.5$.
}
\label{fig1}
\end{figure}

We obtain the dynamical correlation length $L(t)$ in the usual way \cite{Bra94} by monitoring
the one-time space-dependent correlation function 
\begin{equation}
C(t,{\bf r}) = \frac{1}{N} \sum\limits_{\bf{x}}  
\overline{\langle S_{\bf{x}+\bf{r}}(t) S_{\bf{x}}(t) \rangle} ~.
\end{equation}
We define the correlation length $L(t)$ as the distance at which $C(t,{\bf r})$ drops to half of the value it has at ${\bf r}
= {\bf 0}$, see the inset in Fig. \ref{fig1}. The main part of Fig. \ref{fig1} shows in a log-log plot $L$ as a function 
of $t$ for some of the cases studied in this work.
In all cases the initial behavior can be described by a power-law, but deviations from this algebraic
growth already show up when the correlation length is of the order of a few lattice constants. 
These deviations indicate a slowing down of the domain growth after an initial transient behavior.
For example, when plotting for $p = 0.90$
(the case where the deviations are the most notable) $\ln L$ as a function of $\ln(\ln~t)$, one observes that
the data for large times approach a straight line with an effective slope of approximately 2.5 at the end of the run. Even though this
slope is smaller than the value $1/\psi =4$ expected from the work of Huse and Henley \cite{Hus85}, it is
worth stressing that the observed behavior is indeed compatible with a slow crossover to a logarithmic growth
$L(t) \sim \left( \ln t \right)^{1/\psi}$.
Still, much longer times are needed in order to unambiguously characterize this asymptotic regime.

As the power-law increase is
only a transient one, it is not really meaningful to determine an effective dynamical exponent $z$ from this time-dependent length,
as this would only be valid in a certain time window.\cite{foot1} In fact, as we shall see in the following, it is the
naive use of an effective algebraic growth law that is responsible for much of the confusion surrounding the scaling behavior
of disordered systems. Indeed, the correct scaling forms are only revealed when the correct growth law $L(t)$ is properly taken
into account.

\subsection{Autocorrelation function}

\begin{figure}[h]
\centerline{\epsfxsize=5.20in\ \epsfbox{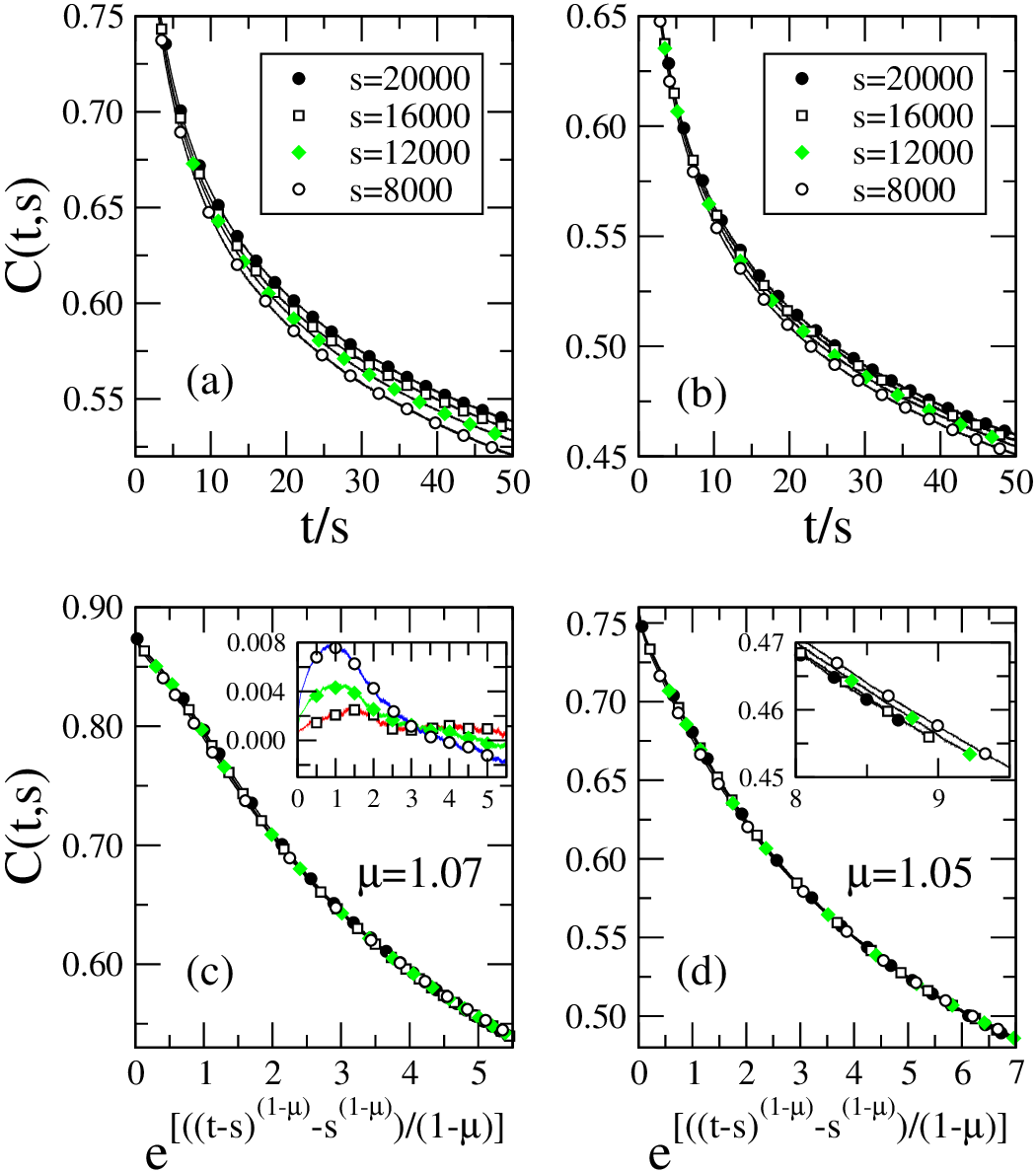}}
\caption{[Color online] Scaling plots of the two-time autocorrelation function for $p = 0.90$, $T = 0.77$ (a,c)  and $p = 0.80$, $T = 0.75$ (b,d).
In traditional full aging plots (where a power-law growth of the correlation length is assumed), see (a) and (b),
data for different waiting times $s$ are plotted as a function of $t/s$.
In the superaging scaling plots,
as proposed in [\onlinecite{Pau07}], see (c) and (d), the fitting parameter $\mu$ is chosen such that the data collapse is optimal
The inset of (c) shows systematic deviations when subtracting
from the s=20000 data the data obtained for s=16000 (red line and open squares), s=12000 (green line and diamonds), or s=8000 (blue line
and open circles).
Choosing $\mu$ such that the data collapse is best for the two largest waiting times, systematic
deviations are observed both for small and for large values of the scaling variable.
This is illustrated in the inset of (d) where we show the waiting time dependent correlation function 
for the largest values of the scaling variable.
Here and in the following error bars are much smaller than the sizes of the symbols.
}
\label{fig2}
\end{figure}

In diluted ferromagnets the scaling properties of the autocorrelation function are strikingly different to those in the
perfect ferromagnets. For the latter systems the data for different waiting times collapse on a single master curve
when plotted as a function of $t/s$,\cite{Hen04} in accordance with the simple aging scaling form (\ref{eq:Cscal}) 
with $b=0$. For the random-site system, however, no data collapse is observed when plotting the two-time
autocorrelation as a function of $t/s$, see Fig. \ref{fig2}a,b. One could try to bring the data to a collapse by letting
$b$ to differ from 0, but this yields the unphysical result that $b < 0$, implying that $C$ would grow without bounds
when $s \longrightarrow \infty$.\cite{Pau07} This observation led Paul {\it et al} \cite{Pau07} to fit their
numerical data to the superaging scaling ansatz (\ref{C_super}). As shown in Fig. \ref{fig2}c,d, this 
fitting ansatz with one free parameter, the exponent $\mu$, indeed yields a data collapse which seems rather 
convincing. A closer look, however, reveals some problems with this ansatz. Choosing $\mu$ such that the collapse is best 
for the two largest waiting times considered, one observes systematic and increasing deviations when going to lower
waiting times. As shown in the inset of Fig. \ref{fig2}c for the case $p = 0.90$ and $T = 0.77$, 
whereas for small values of the scaling variable the data for the 
smaller waiting times are lying below the $s = 20000$ data (the largest waiting
time considered in our study), for larger values of the scaling variable the same data lie above the $s = 20000$ data.
This is also seen in the inset of Fig. \ref{fig2}d where we show for the case $p = 0.80$ and $T = 0.75$
the autocorrelation function for the different waiting times and large values of the scaling variable.
The same behavior being observed for all studied cases, it follows that the superaging scaling ansatz (\ref{C_super})
(which, we recall, is problematic by itself as Kurchan's lemma \cite{Kur02} forbids a superaging
behavior of the autocorrelation)
is a reasonable good fitting function, but that it does not capture the true scaling behavior of the two-time
autocorrelation. 

\begin{figure}[h]
\centerline{\epsfxsize=5.50in\ \epsfbox{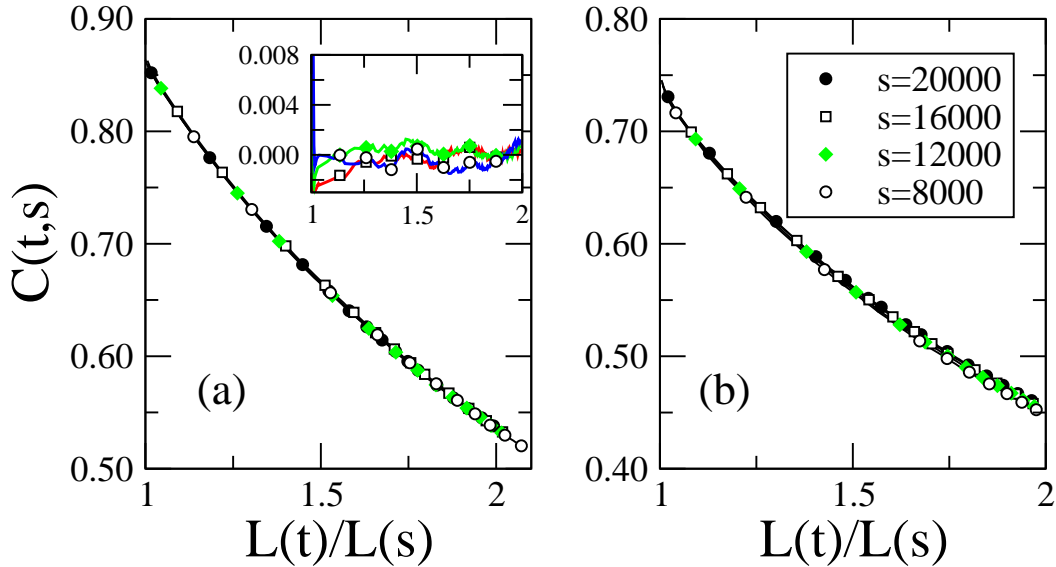}}
\caption{[Color online] Two-time autocorrelation as a function of the ratio $L(t)/L(s)$ where $L(t)$ is the value of the correlation
length at time $t$: (a)  $p = 0.90$, $T = 0.77$ and (b) $p = 0.80$, $T = 0.75$. This parameter-free scaling 
yields a data collapse superior to the superaging scaling, see inset in (a) where 
the data obtained for s=16000 (red line and open squares), s=12000 (green line and diamonds), and s=8000 (blue line and open circles) have been subtracted
from the s=20000 data. In order to facilitate a direct comparison, 
the range of values along the $y$-axis is the same for this inset than
for the inset in Fig. \ref{fig2}c.
}
\label{fig3}
\end{figure}

Faced with this problem, one should remember that the simple aging scaling form (\ref{eq:Cscal}), which 
obviously does not work in Fig. \ref{fig2}a,b, {\it assumes} an algebraic growth of the dynamical correlation
length. For a non-algebraic growth, as we have in our system, see Fig. \ref{fig1}, the correct scaling form  
should be the scaling form (\ref{eq:Cscal2}) where the correlation lengths $L(t)$ and $L(s)$ at times $t$ and
$s$ are used. We test this scaling form in Fig. \ref{fig3} for the same data as in Fig. \ref{fig2} and
indeed find a perfect scaling behavior with $B = 0$. As shown in the inset, no systematic deviations show
up when subtracting from the s=20000 data the data obtained for the smaller waiting times.
Therefore, also in the random ferromagnets a simple aging behavior is observed for the
autocorrelation function, provided that the {\it correct} growth $L(t)$ is taken into account.

Based on this result, we can now give a simple explanation for the origin of 
the apparent differences in the scaling properties of the autocorrelation
in the random-bond and random-site models. Whereas in the random-site model deviations from an algebraic growth law 
already show up when the correlation length is of the order of a few lattice constants, this could
be different for the random-bond model, as various investigations have revealed for that model
an effective algebraic growth over many time decades for most of the studied cases. When both 
the waiting and observation times are in the time window where the growth is effectively algebraic, 
a simple aging scaling $C(t,s) = f(t/s)$ can be
expected.\cite{Hen08} In some case, however, the algebraic power-law growth is presumably not valid during
the whole simulation, which would explain why deviations from the $t/s$ scaling show up.\cite{Hen06} A thorough investigation of
the random-bond Ising model, where the numerically determined length $L(t)$ is used in the scaling, would fully
clarify the situation and help to understand
whether or not there are any fundamental differences between the scaling properties of the random-bond
and the random-site Ising models.

\begin{figure}[h]
\centerline{\epsfxsize=5.90in\ \epsfbox{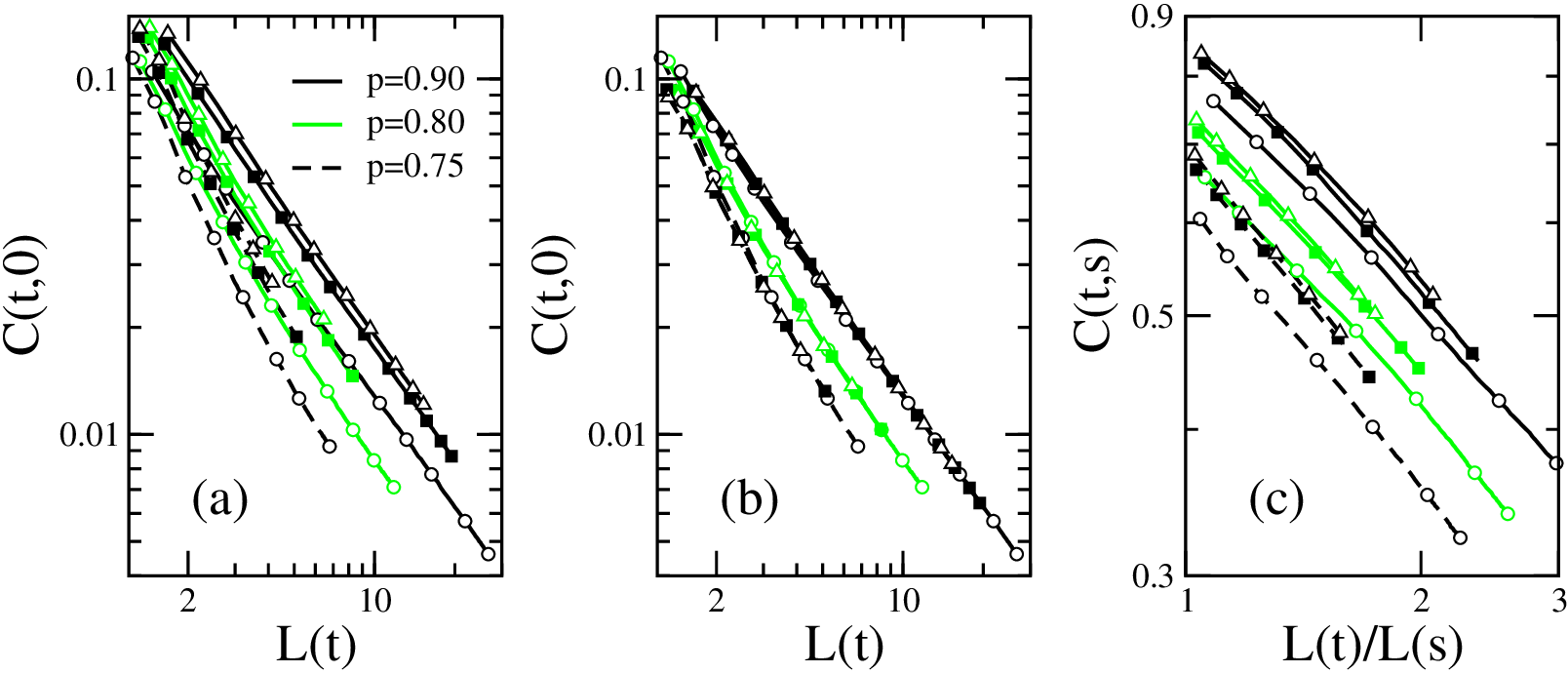}}
\caption{[Color online] (a) $C(t,0)$ as function of $L(t)$ for times up to $t = 20000$. The symbols (open circles: $T= T_c(0.7)$, filled squares: 
$T= T_c(0.5)$, open triangles $T= T_c(0.4)$) are guides for the eyes. In all cases the data rapidly 
display a simple power-law decay. (b) The slopes of $C(t,0)$ are the same for a fixed value of the
dilution $p$ but different temperatures, see Table \ref{table1}. In order to illustrate this, we multiplied
the data for $T= T_c(0.5)$ and $T= T_c(0.4)$ by a constant thus that they fall on the $T= T_c(0.7)$ data for large $L(t)$.
(c) $C(t,s)$ as a function of $L(t)/L(s)$ with s=8000 and $t$ up to $50 s$. The two-time autocorrelation function
has not yet reached the asymptotic power-law regime $L(t) \gg L(s)$ at the end of the run.
}
\label{fig4}
\end{figure}

It is also of interest to determine the autocorrelation exponent that governs the scaling function of the
autocorrelation function for large arguments:
\begin{equation}
C(t,s) \sim \left( \frac{L(t)}{L(s)} \right)^{-\lambda_C}~
\end{equation}
for $L(t)/L(s) \gg 1$. However, in order to reliably measure $\lambda_C$ one usually 
looks at $C(t,0)$ which for long times decays as $\left( L(t) \right)^{-\lambda_C}$. Fig. \ref{fig4}a
shows $C(t,0)= C(L(t))$ for all studied cases in a log-log plot. $C(t,0)$ rapidly approaches a power-law
which makes the measurement of $\lambda_C$ very easy. A simple inspection already reveals that the curves for a given
value of the disorder $p$ decay with similar slopes in  the log-log plot, the slopes being steeper for smaller values of
$p$. In order to make this better visible we have shifted in Fig. \ref{fig4}b the curves for $0.5 T_c(p)$ and $0.4 T_c(p)$
by multiplying $C(t,0)$ by a constant thus that these curves fall on the $0.7 T_c(p)$ curves for larger values of $L(t)$.
The measured values of $\lambda_C$ gathered in Table \ref{table1} indeed show that within error
bars the autocorrelation exponent is independent of the temperature for a fixed value of the dilution. 

It is worth noting that
for all studied cases the value of the autocorrelation exponent is in agreement with the lower bound $d/2$ derived in 
[\onlinecite{Fis88,Yeu96}]. This is in marked contrast to the claim in Ref. [\onlinecite{Pau07}] that this lower bound is violated in
the random-site Ising model. In fact, in their analysis the authors of that paper not only assumed that
$C(t,s) \sim t^{-\lambda_C/z}$, which is not valid in the absence of an algebraic growth law, they also tried
to extract the autocorrelation exponent from $C(t,s)$, which is a notoriously difficult task. This is shown in
Fig. \ref{fig4}c where we plot for the different cases $C(t,s)$ as a function of $L(t)/L(s)$ for $s=8000$ and
$t$ up to $50 s$. One immediately remarks that the values of $L(t)/L(s)$ remain very small and that the
autocorrelation displays a marked curvature for even the largest values of $t$. Obviously, one is not yet in the
scaling regime where $L(t) \gg L(s)$, and a naive measurement at the end of the run would yield effective values
for the autocorrelation exponent that are systematically lower than the asymptotic values readily measured when using
$C(t,0)$.

\begin{table}[tbp]
\begin{center}
\begin{tabular}{|c|c||c|c|c|c|c|c|}
\hline
$p$ & $T$ & $\lambda_C$ & $\lambda_R$ & $A$ \\ \hline\hline
$0.75$ & $0.4 \, T_C(0.75)$ & 1.21(3) & 1.32(3) &  0.30(1)\\ \hline
 & $0.5 \, T_C(0.75)$  & 1.21(2) & 1.31(3) & 0.26(1) \\ \hline
 & $0.7 \, T_C(0.75)$  & 1.18(2) & 1.22(3) & 0.15(2)\\ \hline
$0.80$ & $0.4 \, T_C(0.80)$ & 1.11(1) & 1.14(2) & 0.47(1)\\ \hline
 & $0.5 \, T_C(0.80)$ & 1.10(1) & 1.12(2) & 0.38(1) \\ \hline
 & $0.7 \, T_C(0.80)$ & 1.10(1) & 1.11(2) & 0.30(1) \\ \hline
$0.90$ & $0.4 \, T_C(0.90)$ & 1.06(1) & 1.05(2) & 0.65(1) \\ \hline
 & $0.5 \, T_C(0.90)$ & 1.05(1) & 1.04(2)& 0.63(1) \\ \hline
 & $0.7 \, T_C(0.90)$ & 1.05(1) & 1.06(2) & 0.53(1) \\ \hline
\end{tabular}
\end{center}
\caption{Values of the autocorrelation exponent $\lambda_C$, the
autoresponse exponent $\lambda_R$,  and the scaling exponent $A$ of the response
for all studied cases.
}
\label{table1}
\end{table}

\subsection{Space-time correlation function}

\begin{figure}[h]
\centerline{\epsfxsize=5.50in\ \epsfbox{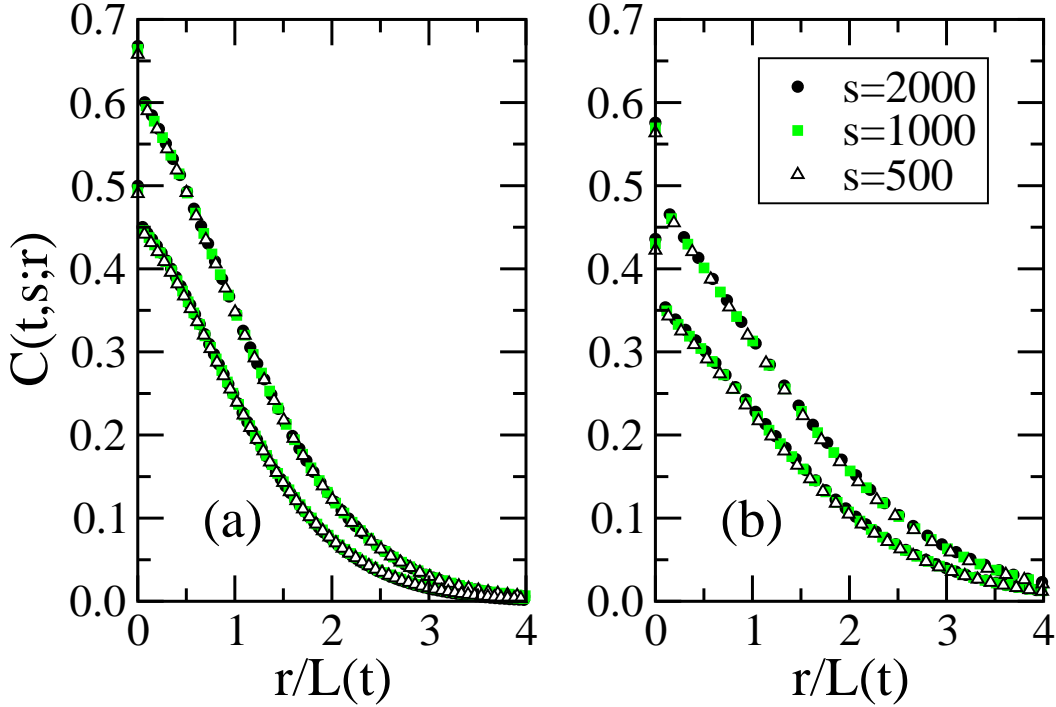}}
\caption{[Color online] Space-time correlation function for various waiting times as a function of $r/L(t)$:
(a) $p=0.90$ and $T = 0.77$, with $L(t)/L(s) = 1.46$ (top) and $L(t)/L(s) = 2.15$ (bottom), and (b)
$p=0.80$ and $T=0.75$, with $L(t)/L(s) = 1.40$ (top) and $L(t)/L(s) = 2.00$ (bottom).
}
\label{fig5}
\end{figure}

It has been noted in the past \cite{Rot06,Hen08} that space- and time-dependent quantities are often better suited
than quantities that only depend on time if one aims at studying the scaling properties of an aging system.
Taking into account what we learned from the autocorrelation, namely that the correct growth law $L(t)$ has to be used
and that the observed simple scaling behavior means that the exponent $B=0$, we should have that the
autocorrelation function $C(t,s,{\bf{r}})$ is only a function of $L(t)/L(s)$ and $r/L(t)$ (or, alternatively, of
$L(t)/L(s)$ and $r/L(s)$):
\begin{equation}
C(t,s,{\bf{r}}) = \widetilde{F}_C\left( \frac{L(t)}{L(s)}, \frac{r}{L(t)} \right)~.
\end{equation}
with $r=\left| \bf{r} \right|$.
We probe this simple scaling in Fig. \ref{fig5} for the cases  $p=0.90$, $T = 0.77$ and $p=0.80$, $T=0.75$ (we obtain 
the same behavior for the other studied cases) where we plot the space-time correlation as a function of the reduced
length $r/L(t)$ for fixed values of $L(t)/L(s)$. The observed data collapse again proves that the correlation function
in the random-site Ising model displays a simple aging scaling behavior.

\begin{figure}[h]
\centerline{\epsfxsize=5.50in\ \epsfbox{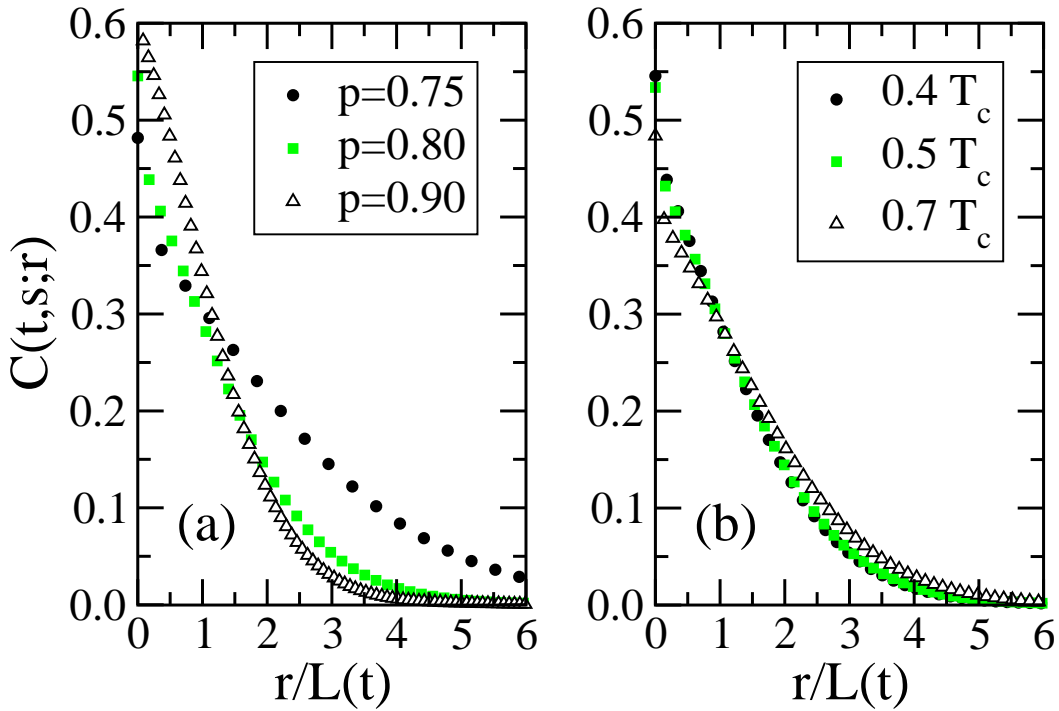}}
\caption{[Color online] (a) Space-time correlation function at $T = 0.4 T_c(p)$ for the different degrees of dilution $p$, 
with $s =1000$ and  $L(t)/L(s) =1.52$.
(b) Space-time correlation function for $p= 0.8$, $s =1000$, and  $L(t)/L(s) =1.52$, for three different
temperatures.
}
\label{fig6}
\end{figure}

At this point one might pause a moment and wonder whether our system has a {\it superuniversal}
scaling behavior. Here superuniversality means that the scaling functions should be independent of 
disorder and temperature once the unique reference length scale has been chosen to be $L(t)$.\cite{Fis88}
This intriguing behavior has indeed been observed in systems with weak randomness
as for example the random-bond Ising model or the random-field Ising model. For our model, the disorder
has a very strong impact on the coarsening process, 
as too much disorder completely destroys phase ordering, and it is a priori not clear whether
a superuniversal behavior is to be expected. As shown in Fig.\ \ref{fig6}a for the temperature
$T = 0.4 T_c(p)$, systems with different dilutions $p$ and fixed value of the ratio $L(t)/L(s)$ do not 
show a common master curve when plotting the space-time correlation as a function of $r/L(t)$. 
This is in agreement with an earlier study by Iwai and Hayakawa \cite{Iwa93} of a diluted system where,
using a cell-dynamical system method, the scaling function of the structure factor was found to depend
on the degree of dilution. Interestingly, we achieve
an approximate data collapse, which gets better the lower the temperatures, when the dilution
is kept fixed and the temperature is changed, see Fig. \ref{fig6}b for the case $p=0.8$. This approximate
superuniversal behavior is in agreement with the expectation that the dynamics for fixed disorder only 
weakly depends on temperature, as long as one is not too close to the critical point.
Still, the observed strong dependence of the scaling function of the space-time correlation on the degree
of dilution indicates that some aspects of the phase ordering of diluted systems remain to be better
understood.

\subsection{Thermoremanent susceptibility}

Finally, we have also extended our study to the thermoremanent susceptibility (\ref{eq:trm}). In order to measure this integrated
response one applies a small magnetic field when quenching the system to the temperature $T$. This field is removed after
the waiting time $s$ and the relaxation to equilibrium is then monitored by measuring the decay of the magnetization.
It is numerically convenient to apply a spatially random magnetic field as a spatially homogeneous field could easily push
the (finite) system into one of the two competing equilibrium states.

\begin{figure}[h]
\centerline{\epsfxsize=5.50in\ \epsfbox{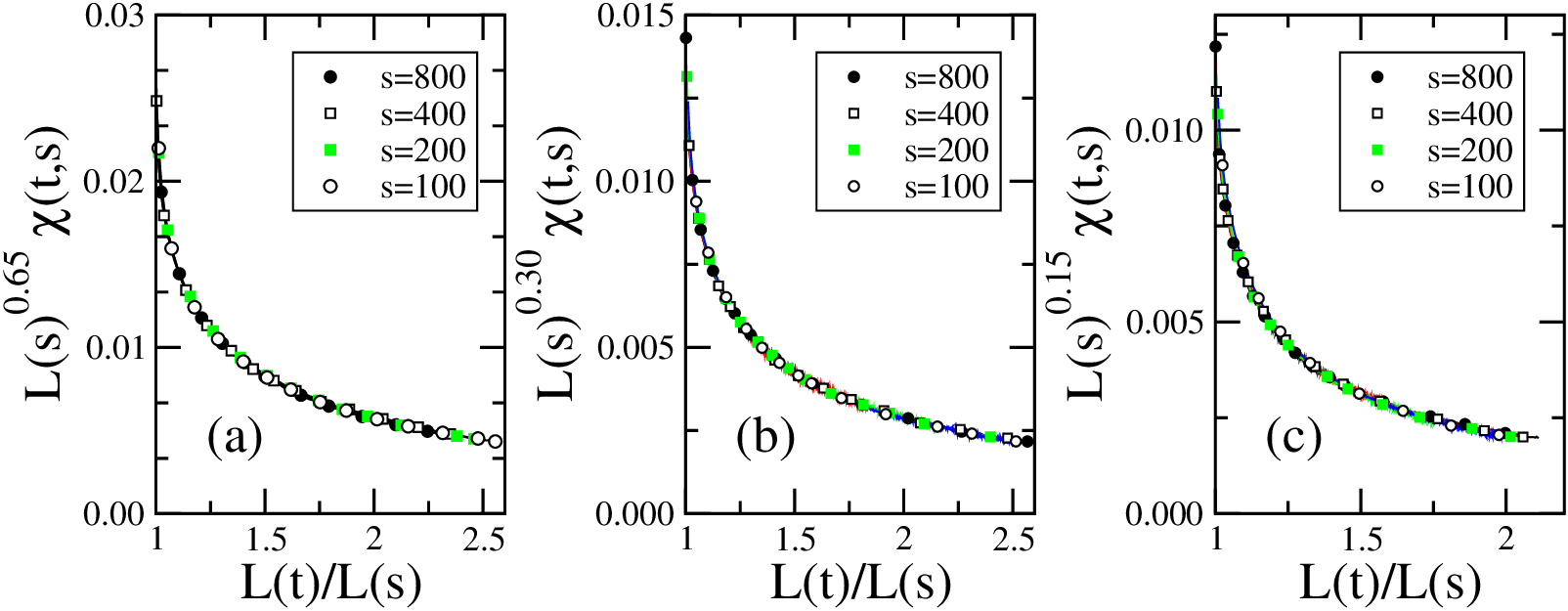}}
\caption{[Color online] Scaling of the thermoremanent susceptibility for (a) 
$p = 0.90$ and $T = 0.77$, (b) $p = 0.80$ and $T = 1.05$, and (c) $p = 0.75$ and $T = 0.91$.
In all cases a simple aging scaling is observed.
}
\label{fig7}
\end{figure}

Figure \ref{fig7} and Table \ref{table1} summarize the observed scaling behavior of the thermoremanent susceptibility.
In all cases we have the simple aging scaling (\ref{eq:trm2}), provided that we use the correct dynamical correlation length $L(t)$.
The non-equilibrium exponent $A$ is thereby found to depend on the dilution and on the temperature. For a given dilution, $A$ increases
when decreasing the temperature, whereas for a fixed value of $T/T_c(p)$ the value of $A$ decreases when decreasing $p$.

As usual the scaling regime is accessed for the response much earlier than for the correlation function. Whereas
for the correlation function finite time corrections start to be negligible for waiting times of s=1000 and larger,
for the integrated response waiting times of $s =100$ and larger already yield a perfect data collapse. This is 
similar to what is observed in the perfect Ising model and in the random-bond Ising model, for example.

We could try to extract from the data shown in Fig. \ref{fig7} the autoresponse exponent $\lambda_R$, 
which governs the power-law decay of the scaling function in the
limit $L(t) \gg L(s)$, but we are confronted with the same problem 
as we encountered when trying to extract the autocorrelation
exponent $\lambda_C$ from the two-time correlation function $C(t,s)$: even for an observation time $t$ that is
fifty times larger than the waiting time $s$, we have that $L(t)$ is only of the order of $2 L(s)$. Consequently, our data still show
a slight curvature in a log-log-plot, and this only allows to obtain a very rough estimate of $\lambda_R$. 
For that reason we determined $\lambda_R$ for the smaller waiting time $s=5$ and $t = 200 s$. 
This allows us to enter the asymptotic power-law regime in most cases. The only exceptions are for $p = 0.75$ and small
temperatures, where the growth is the slowest, as here we still have a slight curvature at the end of the run.
The values of $\lambda_R$ measured in this way are gathered in Table \ref{table1}. These values show the same dependence
on the degree of dilution than $\lambda_C$ and are, in general, consistent with $\lambda_R = \lambda_C$.

Finally, let us briefly comment on the values of the exponent $A$ as this quantity has been of some interest recently.
Thus it has been proposed that $A = 1/2$ for coarsening in two dimensions,\cite{Cor01} and a numerical study of the
random-bond Ising model, that seems to support
this claim, has been published recently.\cite{Lip10} For the random-site Ising model, we can unambiguously conclude that
$A$ is not identical to 1/2. It has been argued in Ref. [\onlinecite{Lip10}] that the thermoremanent magnetization 
(which is proportional to the thermoremanent susceptibility) suffers from
crossover effects that make a reliable estimate of the value of $A$ difficult. However, in our analysis, where we do not
assume a power-law growth but use the correct growth law $L(t)$, we do not have any crossover effects, as seen in Fig. \ref{fig7},
but instead have a very clean scaling behavior with an exponent $A$ that differs from 1/2. 
We find that $A$ depends on temperature and on the degree of dilution, which is similar to what has been
found in an earlier study of the random-bond Ising model.\cite{Hen08} 
Interestingly, systematic deviations from the value 1/2 are also seen in Fig. 5 of Ref. [\onlinecite{Lip10}],
where the exponent $A$ has been determined for the random-bond model
under the assumption of an algebraic growth law. Again, a careful study of
the random-bond case using the correct growth law $L(t)$ should clarify whether there are any qualitative differences between the
random-bond and the random-site models.

\section{Discussion and conclusion}
The law governing the growth of the dynamical correlation length is a crucial quantity in any investigation of
the dynamical scaling in an aging system. In many systems, notably in non-disordered systems, an algebraic growth
rapidly prevails.\cite{HenPle} This is expected to be different in disordered systems as here a crossover from a transient,
pre-asymptotic, power-law regime to an asymptotic regime with a slower, logarithmic, growth is predicted to happen.\cite{Hus85}
There is mounting evidence, due to recent studies of elastic lines in disordered media,\cite{Noh09,Igu09,Mon09}
that this crossover indeed takes place. Obviously, this crossover has to be taken into account in order to elucidate
the scaling properties of disordered systems relaxing towards equilibrium.

In our study of the two-dimensional random-site Ising model, deviations from an algebraic growth law,
compatible with a crossover to a slower, logarithmic growth, indeed show
up on time scales that are accessible in numerical relaxation studies. As a consequence, we do not assume in our scaling analysis
any specific form of the growth law, but directly use the numerically determined dynamical correlation length $L(t)$.
In doing so, a surprisingly simple picture of the relaxation properties of the random-site Ising model emerges.
Indeed, for all the studied quantities (autocorrelation, space-time correlation, and autoresponse function) 
the simple aging scaling behaviors (\ref{eq:Cscal2}), (\ref{eq:Cscal3}), and (\ref{eq:trm2}) are observed when 
plotting the two-times quantities as a function of the ratio of the correlation lengths at times $t$ and $s$: $L(t)/L(s)$.
The autocorrelation has been studied previously in Ref. [\onlinecite{Pau07}] where a superaging scaling ansatz has been used to fit
the numerical data. As we showed in this work, this superaging scaling ansatz, which anyhow is in conflict with very general
theoretical considerations, is hampered by systematic deviations which do not vanish when optimizing the value of
the free parameter $\mu$.

The scaling relations (\ref{eq:Cscal2}), (\ref{eq:Cscal3}), and (\ref{eq:trm2}) are characterized by the values of some 
scaling exponents as well as by the scaling functions themselves. In Table \ref{table1} we report our estimates
for the autocorrelation exponent $\lambda_C$, for the autoresponse exponent $\lambda_R$,
as well as for the exponent $A$ of the response function (in addition, we
found that the value $B=0$ for the exponent of the correlation function, as expected for a phase-ordering system). 
Interestingly, the values of $\lambda_C$ and $\lambda_R$ for a fixed dilution are within error bars independent of the
temperature, with $\lambda_C \approx \lambda_R$. 
This is in agreement with the intuitive picture that the dynamical correlations in the diluted ferromagnets are
mainly governed by the average size and shape of clusters that are occupied by spins. The values
of $\lambda_C$ are always found to be larger than the lower bound $d/2$ derived in
[\onlinecite{Fis88,Yeu96}], similar to what is observed in the random-bond model \cite{Hen08}. 
This corrects an earlier claim \cite{Pau07} that in diluted magnets this bound is violated.
For the exponent $A$, on the other hand, we find that its value depends on the dilution as well as on the temperature.
This is similar to the reported behavior of that exponent in the random-bond Ising model,\cite{Hen08,Lip10} even
though one has to view these results for the random-bond system with some caution as in that analysis a simple
algebraic growth of the correlation length has been assumed. In any case, due to our high quality data and
our careful analysis, we can exclude for the random-site model that the values of $A$ are independent of
dilution and temperature and that they are equal to 1/2.

The attentive reader will have noticed that we did not try to compare the numerically determined scaling functions
with theoretical predictions, as for example those coming from the theory of local scale invariance.\cite{Hen01,Hen02,HenPle}
The reason for that is the non-algebraic growth that prevails in our system. Indeed, the theory of local scale invariance,
which gives explicit predictions for scaling functions in aging systems, predictions that have been found to be valid
in many theoretical and numerical studies,\cite{HenPle} assumes in its present formulation the presence of a unique
length scale that grows as a power-law of time. It is an open and important question whether that theoretical approach
can be generalized to include cases where the growth is non-algebraic.

Besides the disordered magnets, many other disordered systems, as for example vortex glasses in high-temperature 
superconductors,\cite{Bus07} Coulomb glasses,\cite{Shi10} or spin glasses,\cite{Bel09} are undergoing ordering 
processes which have been characterized by an (effective) growth law with a temperature and disorder dependent exponent.
Our results indicate that one has to be very careful in this type of situation as the effective dynamical
exponent presumably only masks the presence of a transient initial time regime, followed by a crossover to a
slower asymptotic growth regime. When studying the dynamical scaling behavior one should not naively assume
an effective growth law, but instead the correct growth law, determined numerically if the exact expression is unknown, 
should be used. A revisitation
of these models and a careful analysis along the lines done here for the random-site Ising model seems 
needed in order to fully clarify the scaling properties of the various disordered systems.

\begin{acknowledgments}
This work was supported by the US Department of Energy
through grant DE-FG02-09ER46613. We thank Malte Henkel for a critical reading of the manuscript.
\end{acknowledgments}


\begin{thebibliography}{99}
\bibitem{foot0} Some systematic deviations, similar to those observed in the random-site model, albeit much less
pronounced, have been reported in an early study of the aging properties of the random-bond model for specific
realizations of the disorder.\cite{Hen06}
\bibitem{foot1} In fact, a non-constant effective dynamical exponent is also seen in Fig. 10(b) of reference,\cite{Pau05} even
though the authors of that paper interpret the data differently.
\end{thebibliography}
\end{document}